\documentclass[%
 reprint,
 superscriptaddress,
 amsmath,amssymb,
 longbibliography,
 prb,
]{revtex4-2}

\usepackage[ruled]{algorithm2e}
\usepackage[noend]{algpseudocode}
\usepackage{multirow}
\usepackage{type1cm} 
\usepackage{graphicx}
\usepackage{dcolumn}
\usepackage{bm}
\DeclareMathAlphabet{\mathpzc}{OT1}{pzc}{m}{it}
\usepackage{color}
\usepackage{float}
\usepackage[usenames,dvipsnames,svgnames,table]{xcolor}
\usepackage{hyperref}
\hypersetup{pdfpagemode=FullScreen,colorlinks=true,linkcolor=NavyBlue,citecolor=BrickRed}

\begin{document}

\title{Low-energy interband Kondo bound states in orbital-selective Mott phases}

\author{Jia-Ming Wang}
\thanks{These authors contributed equally to this work.}
\affiliation{Department of Physics, Renmin University of China, Beijing 100872, China}
\affiliation{Key Laboratory of Quantum State Construction and Manipulation (Ministry of Education), Renmin University of China, Beijing 100872, China}

\author{Yin Chen}
\thanks{These authors contributed equally to this work.}
\affiliation{Department of Physics, Renmin University of China, Beijing 100872, China}
\affiliation{Key Laboratory of Quantum State Construction and Manipulation (Ministry of Education), Renmin University of China, Beijing 100872, China}

\author{Yi-Heng Tian}
\thanks{These authors contributed equally to this work.}
\affiliation{Department of Physics, Renmin University of China, Beijing 100872, China}
\affiliation{Key Laboratory of Quantum State Construction and Manipulation (Ministry of Education), Renmin University of China, Beijing 100872, China}

\author{Rong-Qiang He}\email{rqhe@ruc.edu.cn}
\affiliation{Department of Physics, Renmin University of China, Beijing 100872, China}
\affiliation{Key Laboratory of Quantum State Construction and Manipulation (Ministry of Education), Renmin University of China, Beijing 100872, China}

\author{Zhong-Yi Lu}\email{zlu@ruc.edu.cn}
\affiliation{Department of Physics, Renmin University of China, Beijing 100872, China}
\affiliation{Key Laboratory of Quantum State Construction and Manipulation (Ministry of Education), Renmin University of China, Beijing 100872, China}
\affiliation{Hefei National Laboratory, Hefei 230088, China}

\date{\today}

\begin{abstract}
Low-energy excitations in correlated electron systems may show intricate behaviors and provide essential insights into the dynamics of quantum states and phase transitions. Here, we study a typical half-filled two-orbital Hubbard model featuring the so-called holon-doublon (HD) low-energy excitations in the orbital-selective Mott phase (OSMP), where the principal form of the low-energy excitations has been considered to be a HD bound state. We employ standard single-site dynamical mean-field theory (DMFT), using NORG as an improved impurity solver to calculate the spectral functions at zero temperature. We show that the HD bound state gives an incomplete or even wrong picture for the low-energy excitations. Instead, the excitations are composed of a Kondo-like state in the wide band and a doublon in the narrow band, termed as interband Kondo-like (IBK) bound states. Remarkably, we find that, as the bandwidths of the two bands approach each other, anomalous IBK bound-state excitations appear in the metallic {\em wide} band. Our study provides a new picture for the low-energy excitations in the OSMP.
\end{abstract}
\maketitle

\section{Introduction}
The Mott transition, which indicates the transformation of a quantum many-body system from metallic to insulating states driven by strong electron interactions, is a pivotal phenomenon in condensed matter physics \cite{Mott1968rmp, Imada1998rmp}. In multi-orbital systems with varying bandwidths, the phase in which some bands are Mott insulating while the others remain metallic is known as an orbital-selective Mott phase (OSMP) \cite{Koga2004prl, Liebsch2005prl}. This phenomenon was initially discussed widely in theoretical studies \cite{Koga2005prb, Ferrero2005prb, Medici2009prl, Medici2011prb, Ouyang2024arxiv}. Recently, it has been found experimentally in some materials \cite{Vojta2010jltp, Zhu2022prb, Kim2022prb, Huang2022ncp, Cuono2023sci, Horio2023cp, Kim2024prl}.

In earlier years, research on low-energy excitations near Mott transitions was limited to single-band Hubbard models \cite{Ziqiang2014prb, Andreas2017prl}. Recently, N\'u\~nez-Fern\'andez \textit{et al.} \cite{Nunez2018prb} studied a half-filled two-orbital Hubbard model as a typical representative of multi-orbital Hubbard models but being the simplest,
\begin{align}\label{eq:lattmodel}
  H &= -\sum_{\langle ij\rangle, l, \sigma} t_{l} c_{il\sigma}^{\dagger} c_{jl\sigma} + U \sum_{i,l} (n_{il\uparrow} - \frac{1}{2}) (n_{il\downarrow} - \frac{1}{2}) \nonumber \\
    &\quad + U^\prime \sum_{i} (n_{i1} - 1) (n_{i2} - 1),
\end{align}
where $c_{il\sigma}^{\dagger}$ and $c_{il\sigma}$ are the electron creation and annihilation operators for orbital $l$ on site $i$ with spin $\sigma$, $U$ ($U^\prime$) is the intra- (inter-)orbital Coulomb repulsion, $n_{il\sigma}=c_{il\sigma}^{\dagger}c_{il\sigma}$, $n_{il}=\sum_{\sigma}n_{il\sigma}$, and $\langle ij\rangle$ indicates that only the nearest-neighbor hoppings are considered. $l = 1, 2$ denote the wide band (WB) and narrow band (NB), respectively, namely $t_1 > t_2$. As $U$ increases, with $\Delta = U - U^\prime > 0$, the NB first transitions into a Mott insulating state, while the WB remains metallic, indicating that the system has entered an OSMP. This is illustrated in Fig.~\ref{fig:Fig1}(a). N\'u\~nez-Fern\'andez \textit{et al.} \cite{Nunez2018prb} identified new quasiparticle (QP) states. In the OSMP, these QP peaks persist as in-gap states around $\omega = \pm \Delta$ in the Mott gap of the insulating NB. They also introduced the concept of the holon-doublon (HD) bound state, in which a doublon in the NB binds with a holon in the WB. They proposed that the QPs are mainly formed by the HD bound states at $\omega = \pm \Delta$. At the symmetric point $\Delta = 0$ ($U = U'$), the QP peaks are located at the Fermi energy, leading to a simultaneous Mott transition in the two band even if the bandwidths are different.

The discovery of the HD bound state inspired subsequent research on low-energy excitations in two-orbital systems. Niu {\it et al.} \cite{Niu2019prb} studied a Kanamori model, which extends the two-orbital Hubbard model (\ref{eq:lattmodel}) by adding spin-flip and pair-hopping Hund interactions, and found similar QP low-energy excitations. Hallberg {\it et al.} \cite{Hallberg2020prb} studied a hole-doped system and observed that the Hund's coupling splits the HD excitation peaks. Subsequent studies extending to three-orbital systems \cite{Sroda2023prb} as well as using different numerical methods, such as the slave-spin method \cite{Komijani2019prb}, numerical renormalization group \cite{Kugler2019prb}, and density matrix renormalization group \cite{Boidi2021prr, Aucar2024prb}, have also found similar excitations. Thus far, comprehensive theoretical studies of the low-energy excitations in the OSMP agree that the HD bound states are a principal form of the low-energy excitations for multi-orbital Hubbard models \cite{Nunez2018prb, Niu2019prb, Komijani2019prb, Kugler2019prb, Hallberg2020prb, Boidi2021prr, Sroda2023prb, Aucar2024prb}. However, our study shows that this picture is incomplete or even wrong.

In this paper, we employ standard single-site dynamical mean-field theory (DMFT) method with the natural orbitals renormalization group (NORG) \cite{He2014prb, He2015prb, Zheng2018cpl} as the impurity solver to study the low-energy excitation behavior of the two-orbital model (\ref{eq:lattmodel}).
By calculating the density of states (DOS) and special Green's functions, we find previously undetected components of the low-energy excitations in the Mott insulating narrow band. These components reveal a new picture for the low-energy excitations: they are an interband bound state composed of a Kondo-like QP state in the WB and a doublon/holon in the NB. We term these low-energy excitation states as interband Kondo-like (IBK) bound states. Additionally, when the bandwidths of the two bands approach each other, we have observed two anomalous IBK bound states around $\omega = \pm 2\Delta$ in the WB.


\section{Model and method}

We consider the half-filled two-orbital Hubbard model, described by Eq.~(\ref{eq:lattmodel}), on a Bethe lattice with an infinite coordination number \cite{Georges1996rmp}, resulting in a half-bandwidth of $D_l = 2t_l$. We set $t_1 = 0.5$ so that the WB's half-bandwidth $D_1 = 1$, which we adopt as the energy unit. We utilize DMFT with NORG to study this model. For technical details, see Appendices A, B, and C.

\begin{figure}[b] 
  \includegraphics[width=8.6cm]{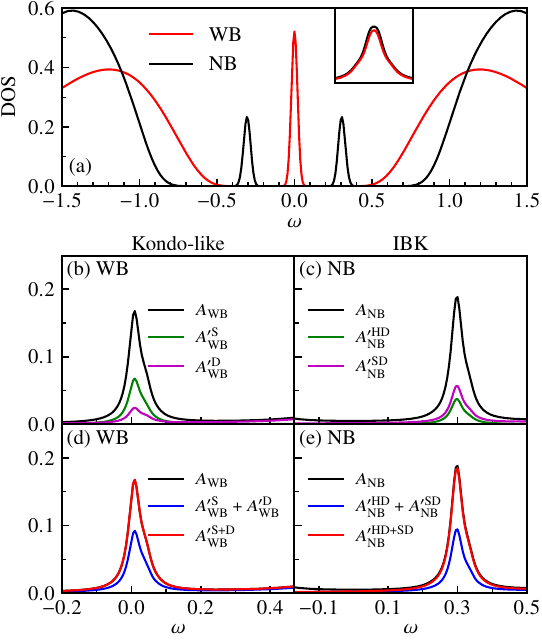}
  \caption{
    (a) DOS for the two-orbital model with $U = 3$, $\Delta = 0.3$, and $t_2 = 0.1t_1$. In the WB, a prominent Kondo-like QP peak is evident around the Fermi level, and in the NB, two distinct QP peaks are visible around $\omega = \pm \Delta$. Inset: the total spectral weight of the two peaks around $\omega = \pm \Delta$ in the NB matches the spectral weight of the Kondo-like QP peak in the WB. 
    (b-e) Unraveling the excitation spectra in the WB and NB and comparing with the standard Green's function $A \equiv -\frac{1}{\pi} \mathrm{Im} G^{>}$.
    Individual special Green's functions (b) $A^{\prime \rm {S}}_{\rm {WB}}$ and $A^{\prime \rm {D}}_{\rm {WB}}$, (c) $A^{\prime \rm {HD}}_{\rm {NB}}$ and $A^{\prime \rm {SD}}_{\rm {NB}}$. 
    Combined special Green's functions (d) $A^{\prime \rm {S}}_{\rm {WB}} + A^{\prime \rm {D}}_{\rm {WB}}$ and $A^{\prime \rm {S+D}}_{\rm {WB}}$, (e) $A^{\prime \rm {HD}}_{\rm {NB}} + A^{\prime \rm {SD}}_{\rm {NB}}$ and $A^{\prime \rm {HD+SD}}_{\rm {NB}}$. See the text for details.
    }\label{fig:Fig1}
\end{figure}

In Fig.~\ref{fig:Fig1}(a), we present the DOS, demonstrating that the NB becomes insulating while the WB remains metallic, indicating that the system enters an OSMP under certain conditions ($U = 3$, $\Delta = 0.3$, and $t_2 = 0.1t_1$). Notably, the WB exhibits a Kondo-like QP peak around $\omega = 0$, while the NB has two low-energy QP peaks around $\omega = \pm \Delta$.

To further clarify the nature of the low-energy excitations, we define and calculate the special Green's functions to unravel the spectral function:
\begin{align}
  A^\prime(\omega) &\equiv -\frac{1}{\pi} \mathrm{Im} G^{\prime >}(\omega) \nonumber\\
  &= -\frac{1}{\pi} \mathrm{Im} \left\langle \phi \middle| (\omega + i \eta - H_{\mathrm{imp}})^{-1} \middle| \phi \right\rangle,
  \label{eq:special_green}
\end{align}
where $\left| \phi \right\rangle$ is a single-particle excitation state and $H_{\mathrm{imp}}$ is the Hamiltonian for the DMFT-mapped quantum impurity model (for details, see Appendix A). For the excitations in the NB, $\left| \phi \right\rangle$ can be one of the following states:
\begin{equation}
  \begin{cases}
    \left| \phi^{\rm HD}_{\rm NB} \right\rangle &=(1-n_{1\uparrow})(1-n_{1\downarrow})n_{2\downarrow}c_{2\uparrow}^{\dagger}\left| \text{gs} \right\rangle,   \\
    \left| \phi^{\rm SD}_{\rm NB} \right\rangle &=n_{1\uparrow}(1-n_{1\downarrow})n_{2\downarrow}c_{2\uparrow}^{\dagger}\left| \text{gs} \right\rangle \\&+ (1-n_{1\uparrow})n_{1\downarrow}n_{2\downarrow}c_{2\uparrow}^{\dagger}\left| \text{gs} \right\rangle,   \\
    \left| \phi^{\rm DD}_{\rm NB} \right\rangle &=n_{1\uparrow}n_{1\downarrow}n_{2\downarrow}c_{2\uparrow}^{\dagger}\left| \text{gs} \right\rangle, 
  \end{cases} \label{eq:nb-ex_state}
\end{equation}
where $\left| \text{gs} \right\rangle$ refers to the ground state of the impurity model, $c_{l\sigma}^{\dagger}$ is the electron creation operator for the impurity orbital $l$ with spin $\sigma$, and $n_{l\sigma} = c_{l\sigma}^{\dagger} c_{l\sigma}$. The impurity orbital of the NB, after inserting a particle, becomes mainly doubly occupied, and the impurity orbital of the WB may be empty ($\left| \phi_{\rm NB}^{\rm HD} \right\rangle$), singly occupied ($\left| \phi_{\rm NB}^{\rm SD} \right\rangle$), or doubly occupied ($\left| \phi_{\rm NB}^{\rm DD} \right\rangle$). Noting that $\left| \phi_{\rm NB}^{\rm HD} \right\rangle + \left| \phi_{\rm NB}^{\rm SD} \right\rangle + \left| \phi_{\rm NB}^{\rm DD} \right\rangle = n_{2\downarrow}c_{2\uparrow}^{\dagger}\left| \text{gs} \right\rangle$ with a doublon in the NB. Note that the system is particle-hole symmetric and our analysis focuses only on the particle excitation case.


\section{Results and discussion}
\subsection{Kondo-like excitations}
The WB exhibits a Kondo-like QP peak around $\omega = 0$, as shown in Fig.~\ref{fig:Fig1}(a). We unravel the DOS for the WB using Eq.~(\ref{eq:special_green}). In Fig.~\ref{fig:Fig1}(b), $A_{\rm WB}$ represents the standard Green's function, $A_{\rm WB} \equiv -\frac{1}{\pi} \mathrm{Im} G^{>}_{\rm WB}$. $A'^{\rm S}_{\rm WB}$ and $A'^{\rm D}_{\rm WB}$ are the special Green's functions that correspond to the particle excitation states with single occupancy $\left| \phi^{\rm S}_{\rm WB} \right\rangle = (1-n_{1\downarrow})c_{1\uparrow}^{\dagger}\left| \text{gs} \right\rangle$ and double occupancy $\left| \phi^{\rm D}_{\rm WB} \right\rangle = n_{1\downarrow}c_{1\uparrow}^{\dagger}\left| \text{gs} \right\rangle$ in the WB, respectively. We have observed that both $A'^{\rm S}_{\rm WB}$ and $A'^{\rm D}_{\rm WB}$ are nonzero around $\omega = 0$. This indicates that the particle excitation state has contributions from both the $\left| \phi^{\rm S}_{\rm WB} \right\rangle$ and $\left| \phi^{\rm D}_{\rm WB} \right\rangle$ components. In Fig.~\ref{fig:Fig1}(d), we introduce a special Green's function $A'^{\rm S + D}_{\rm WB}$ with $\left| \phi \right\rangle = \left| \phi^{\rm S}_{\rm WB} \right\rangle + \left| \phi^{\rm D}_{\rm WB} \right\rangle$ in Eq.~(\ref{eq:special_green}) and compare it with $A_{\rm WB}$ and the sum $A'^{\rm S}_{\rm WB} + A'^{\rm D}_{\rm WB}$. We find that $A'^{\rm S + D}_{\rm WB}$ matches $A_{\rm WB}$, while $A'^{\rm S}_{\rm WB} + A'^{\rm D}_{\rm WB}$ does not. The difference between the two is that the off-diagonal parts, $\left\langle \phi^{\rm S}_{\rm WB} \middle| (\omega + i \eta - H_{\mathrm{imp}})^{-1} \middle| \phi^{\rm D}_{\rm WB} \right\rangle$ and $\left\langle \phi^{\rm D}_{\rm WB} \middle| (\omega + i \eta - H_{\mathrm{imp}})^{-1} \middle| \phi^{\rm S}_{\rm WB} \right\rangle$, are present in the former but missing in the latter. This shows that the Kondo-like particle excitation state is a superposition state of the singly and doubly occupied states in the WB.

\begin{figure}[b] 
  \includegraphics[width=8.6cm]{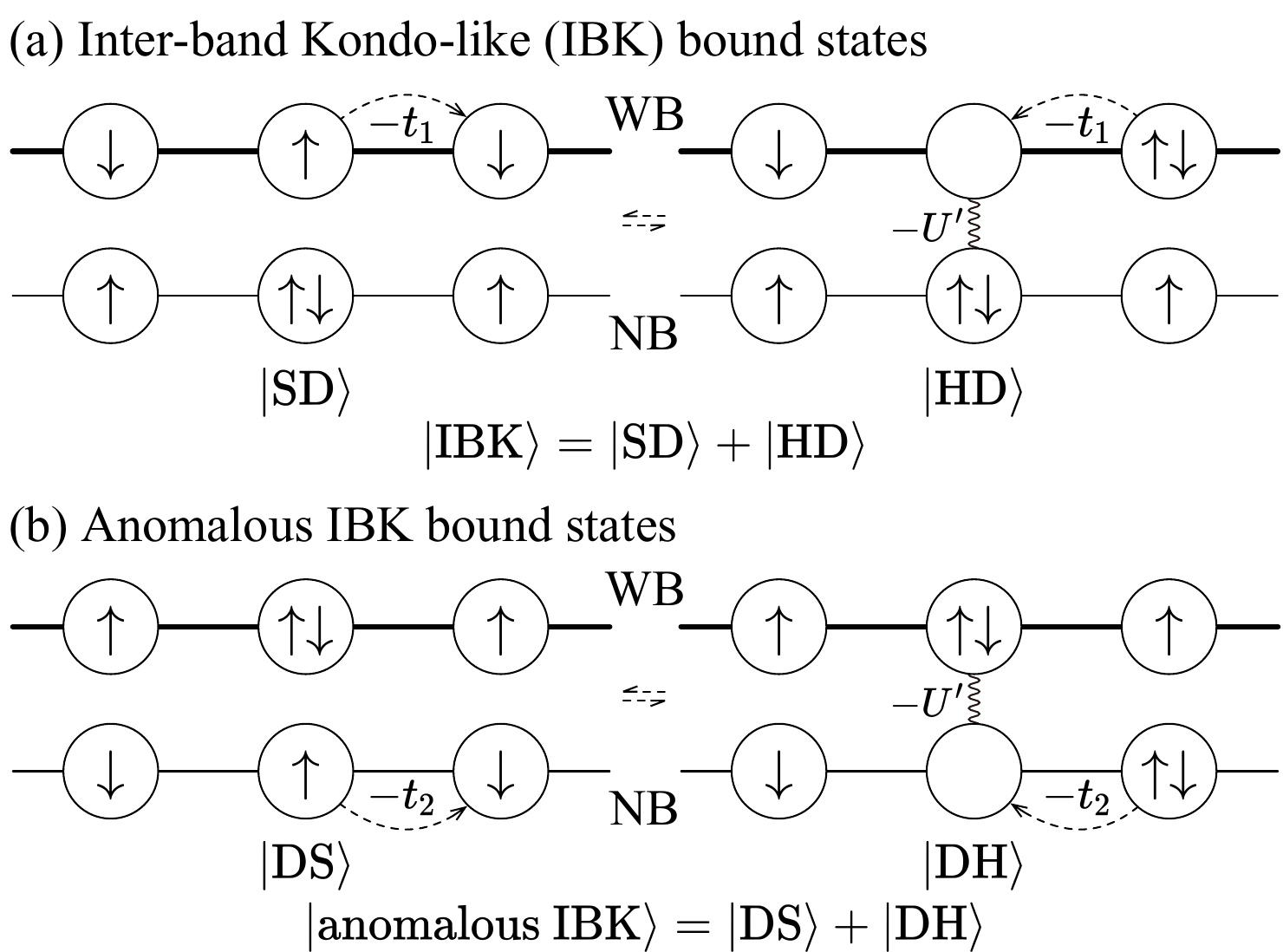}
  \caption{
    Schematic diagrams of the IBK bound states (a) and the anomalous IBK bound states (b). The WB are represented by thicker lines, while the NB are depicted with thinner lines, indicating their respective hopping integrals $-t_1$ for the WB and $-t_2$ for the NB. For the (anomalous) IBK bound states, the existence of an HD (DH) component allows the system to reduce its $U^\prime$ energy cost. The double arrow indicates that the SD and HD (DS and DH) states form superposition states.
    }\label{fig:Fig2}
\end{figure}

\subsection{IBK bound states}
We show the calculated special Green's functions of the NB (Eq.~(\ref{eq:special_green})) in Fig.~\ref{fig:Fig1}(c). Surprisingly, in addition to an HD component around $\omega = \Delta$, an SD component is also identified. This shows that the HD component alone is insufficient to account for the low-energy excitation. Thus, this low-energy excitation is not an HD bound state (with a localized holon in the WB and a localized doublon in the NB). To further unravel this excitation, we introduce a special Green's function $A^{\prime {\rm HD} + {\rm SD}}_{\rm NB}$ with $\left| \phi \right\rangle = \left| \phi^{\rm HD}_{\rm NB} \right\rangle + \left| \phi^{\rm SD}_{\rm NB} \right\rangle$ in Eq.~(\ref{eq:special_green}) and compare it with $A^{\prime \rm HD}_{\rm NB} + A^{\prime \rm SD}_{\rm NB}$ and $A_{\rm NB} \equiv -\frac{1}{\pi} \mathrm{Im} G^{>}_{\rm NB}$ in Fig.~\ref{fig:Fig1}(e). $A^{\prime {\rm HD} + {\rm SD}}_{\rm NB}$ perfectly matches $A_{\rm NB}$, while the sum $A^{\prime \rm HD}_{\rm NB} + A^{\prime \rm SD}_{\rm NB}$ does not. The difference between the two is that the off-diagonal parts, $\left\langle \phi_{\rm NB}^{\rm SD} \middle| (\omega + i \eta - H_{\mathrm{imp}})^{-1} \middle| \phi_{\rm NB}^{\rm HD} \right\rangle$ and $\left\langle \phi_{\rm NB}^{\rm HD} \middle| (\omega + i \eta - H_{\mathrm{imp}})^{-1} \middle| \phi_{\rm NB}^{\rm SD} \right\rangle$, are present in the former but missing in the latter. Therefore, in the low-energy excitation state, the impurity orbital of the WB is neither in a localized holon state nor in a localized single-occupancy state, but rather in a superposition state involving both the impurity and the bath, noting that the WB itself is particle-number conserved. It turns out that the NB is in a localized doublon state, while the WB is not in a localized state involving only the impurity orbital. Additionally, the inset of Fig.~\ref{fig:Fig1}(a) shows that the total spectral weight of the two QP peaks around $\omega = \pm \Delta$ in the NB is approximately equal to the spectral weight of the Kondo-like QP peak around the Fermi level. We term the excitation states around $\omega = \pm \Delta$ in the NB as interband Kondo-like (IBK) bound states.

We use a Bethe lattice with a coordination number of two to further explain the key features of an IBK bound state in the OSMP, with a schematic of this state shown in Fig.~\ref{fig:Fig2}(a). Inserting an electron into the orbital at the middle site in the NB leads to a doublon in the NB. An HD component is created when the doublon binds with an empty state (a holon) in the metallic WB, as depicted in the right part of Fig.~\ref{fig:Fig2}(a). The interaction potential energy of the HD component is only $\Delta$ higher than that of the ground state but lower than the interaction potential energies of the SD and the DD components. More specifically, as described in Eq.~(\ref{eq:lattmodel}), the interaction potential energy for the middle orbital of the HD component is $\frac{1}{4} U$ for both the wide and narrow bands, and $-U^\prime$ for the inter-orbital part. Comparing this to the ground-state potential energy of the middle site, which is $-\frac{1}{4} U$ because of single occupancy in both the WB and NB, one finds that the energy gap between the HD component and the ground state is $\Delta = U - U^\prime$. In contrast, the energy gap between the DD (SD) component and the ground state is $U + U^\prime$ ($\frac{1}{2} U$). Noting that the DD component is much higher than $\Delta$, this excludes the DD component from the low-energy excitation states.

Since the WB is close to the Mott transition, the kinetic energy in the WB is comparable to the potential energy. This makes the empty state in the orbital at the middle site of the WB, namely the holon in the WB, unstable. This instability allows an electron to appear at the middle site in the WB, leading to the existence of an SD component, as depicted in the left part of Fig.~\ref{fig:Fig2}(a). The nonzero off-diagonal contributions (e.g., $\left\langle \phi_{\rm NB}^{\rm SD} \middle| (\omega + i \eta - H_{\mathrm{imp}})^{-1} \middle| \phi_{\rm NB}^{\rm HD} \right\rangle \neq 0$) demonstrate that an IBK bound state is {\em a superposition of empty and singly occupied states in the WB} with a doublon in the NB, as marked by the middle double arrow in Fig.~\ref{fig:Fig2}(a). The interplay between intra- and inter-orbital interactions results in the IBK bound states being $\Delta$ higher in energy than the ground state. Note that the Kondo-like state in the WB is not produced by a single-particle excitation in the WB but by a single-particle excitation in the NB. For this excitation state, the interband interaction $U'$ plays a key role.

\begin{figure}[t]
  \includegraphics[width=8.6cm]{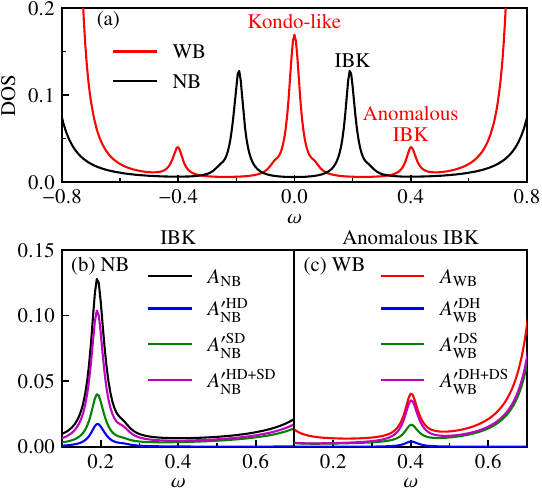} 
  \caption{
    (a) DOS for the two-orbital model with $U = 3.3$, $\Delta = 0.2$, and $t_2 = 0.8 t_1$. 
    (b-c) Unraveling the excitation spectra in the NB and WB and comparing with the standard Green's function $A \equiv -\frac{1}{\pi} \mathrm{Im} G^{>}$.
    (b) NB's special Green's functions $A^{\prime \mathrm{HD}}_{\mathrm{NB}}$, $A^{\prime \mathrm{SD}}_{\mathrm{NB}}$, and $A^{\prime \mathrm{HD+SD}}_{\mathrm{NB}}$. 
    (c) WB's special Green's functions $A^{\prime \mathrm{DH}}_{\mathrm{WB}}$, $A^{\prime \mathrm{DS}}_{\mathrm{WB}}$, and $A^{\prime \mathrm{DH+DS}}_{\mathrm{WB}}$.  See the text for details.
    }\label{fig:Fig3}
\end{figure}

\subsection{Anomalous IBK bound states}

We consider another scenario where the bandwidths of the two bands approach each other. In Fig.~\ref{fig:Fig3}(a), we present the DOS with $t_2 = 0.8 t_1$, $U = 3.3$, and $\Delta = 0.2$. The WB exhibits a Kondo-like QP peak around $\omega = 0$, while the NB has two IBK QP peaks around $\omega = \pm \Delta$, similar to what we discuss above for the normal IBK bound state. Unexpectedly, in Fig.~\ref{fig:Fig3}(a), we have observed additional QP peaks around $\omega = \pm 2\Delta$ in the WB, distinct from the Kondo-like QP peak around the Fermi level.

\begin{figure}[t] 
  \includegraphics[width=8.6cm]{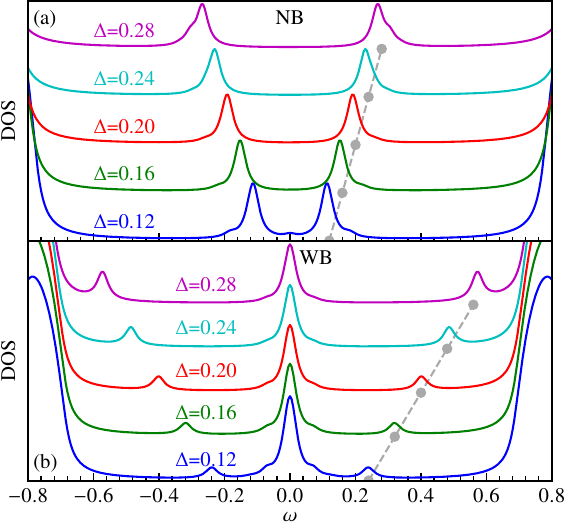} 
  \caption{
    Dependence of the IBK QP peaks on $\Delta$. $U = 3.3$, $t_2 = 0.8 t_1$. $\Delta =$ 0.12, 0.16, 0.20, 0.24, and 0.28. The IBK excitation peaks around $\omega = \pm \Delta$ in the NB (a) and around $\omega = \pm 2\Delta$ in the WB (b) are clearly seen. The grey dashed lines guide the eye for the positions of the QP peaks.
    }\label{fig:Fig4}
\end{figure}

To confirm these additional QP peaks, we calculate the DOS with different $\Delta$ while keeping other parameters fixed. In Fig.~\ref{fig:Fig4}, we show the dependence of the QP peaks with $\Delta$ in both the NB and WB. The IBK bound states in the NB are observed around $\omega = \pm \Delta$, as shown in Fig.~\ref{fig:Fig4}(a). In Fig.~\ref{fig:Fig4}(b), besides the Kondo-like QP peak, we observe additional QP peaks in the WB around $\omega = \pm 2\Delta$.

To understand the additional excitation peaks at $\omega = \pm 2\Delta$, we applied a similar analysis used for the IBK bound states in the NB and calculated the special Green's functions $A_{\rm WB}$, $A^{\prime {\rm DH}}_{\rm WB}$, $A^{\prime {\rm DS}}_{\rm WB}$, and $A^{\prime {\rm DH+DS}}_{\rm WB}$ as described in Eq.~(\ref{eq:special_green}). The impurity orbital of the WB, after inserting a particle, becomes mainly doubly occupied. The impurity orbital of the NB may be empty ($\left| \phi_{\rm WB}^{\rm DH} \right\rangle = (1-n_{2\uparrow})(1-n_{2\downarrow})n_{1\downarrow}c_{1\uparrow}^{\dagger}\left| \text{gs} \right\rangle$), singly occupied ($\left| \phi_{\rm WB}^{\rm DS} \right\rangle = n_{2\uparrow}(1-n_{2\downarrow})n_{1\downarrow}c_{1\uparrow}^{\dagger}\left| \text{gs} \right\rangle + (1-n_{2\uparrow})n_{2\downarrow}n_{1\downarrow}c_{1\uparrow}^{\dagger}\left| \text{gs} \right\rangle$), or doubly occupied ($\left| \phi_{\rm WB}^{\rm DD} \right\rangle = n_{2\uparrow}n_{2\downarrow}n_{1\downarrow}c_{1\uparrow}^{\dagger}\left| \text{gs} \right\rangle$). Note that $\left| \phi_{\rm WB}^{\rm DH} \right\rangle + \left| \phi_{\rm WB}^{\rm DS} \right\rangle + \left| \phi_{\rm WB}^{\rm DD} \right\rangle = n_{1\downarrow}c_{1\uparrow}^{\dagger}\left| \text{gs} \right\rangle$ with a doublon in the WB. We compare these special Green's functions with the standard Green's function $A_{\rm WB}$. As shown in Fig.~\ref{fig:Fig3}(c), both $A^{\prime {\rm DH}}_{\rm WB}$ and $A^{\prime {\rm DS}}_{\rm WB}$ are nonzero at $\omega = 2\Delta$. The $A^{\prime {\rm DH+DS}}_{\rm WB}$ matches $A_{\rm WB}$, while both $A^{\prime {\rm DH}}_{\rm WB}$ and $A^{\prime {\rm DS}}_{\rm WB}$ are smaller than $A_{\rm WB}$. This indicates that the excitation state has off-diagonal contributions from both DS and DH components. Therefore, the impurity orbital of the NB is neither in a localized single-occupancy state nor in a localized holon state. Instead, it is in a superposition state involving both empty and singly occupied states, reminiscent of the IBK bound state in the NB discussed above. However, these states are quite anomalous, leading localized insulating NB states to have Kondo-like excitation states, which contradicts the typical characteristics of the localized Mott states. This shows that the particle excitation in the WB can affect the NB's localization and insulation in the anomalous IBK bound states.

To further explain the anomalous features of IBK bound states in the WB around $\omega = \pm 2\Delta$, we show a schematic of these states in Fig.~\ref{fig:Fig2}(b). Inserting an electron into the orbital at the middle site in the WB leads to a doublon in the WB. An DH component is created when the doublon binds with an empty state (a holon) in the insulating NB, as depicted in the right part of Fig.~\ref{fig:Fig2}(b). The DH component gains an advantage in interaction potential energy for reasons similar to the HD component (discussed in normal IBK bound states in the NB). Since the NB is close to the Mott transition, the kinetic energy in the NB is comparable to the potential energy. This makes the empty state in the orbital at the middle site of the NB, namely the holon, unstable. This instability allows an electron to appear at the middle site in the NB, leading to the existence of an DS component, as depicted in the left part of Fig.~\ref{fig:Fig2}(b). The nonzero off-diagonal contributions (e.g., $\left\langle \phi_{\rm WB}^{\rm DS} \middle| (\omega + i \eta - H_{\mathrm{imp}})^{-1} \middle| \phi_{\rm WB}^{\rm DH} \right\rangle \neq 0$) demonstrate that an anomalous IBK bound state is {\em a superposition of empty and singly occupied states in the NB} with a doublon in the WB, as marked by the middle double arrow in Fig.~\ref{fig:Fig2}(b). However, the position of the anomalous IBK QP peaks (around $\omega = \pm 2\Delta$) in the WB cannot be explained with the interaction potential energy, unlike the normal IBK excitations in the NB.


\section{Conclusion}

Using state-of-the-art numerical calculations for the two-orbital Hubbard model with a repulsive inter-orbital interaction, we have found that the bound-state excitations appearing in the Mott gap of the NB in the OSMP is composed of a Kondo-like QP state in the WB and a doublon in the NB. This corrects the view of the HD bound state in the literature. When the bandwidths of the two bands approach each other, we have found anomalous IBK bound states around $\omega = \pm 2\Delta$ {\em in the WB}. This discovery challenges the conventional understanding that the Mott insulating bands decouple from the other bands in an OSMP.

\begin{acknowledgments}
    This work was supported by National Natural Science Foundation of China (Grant No. 11934020). Z.Y.L. was also supported by Innovation Program for Quantum Science and Technology (Grant No. 2021ZD0302402). J.M.W. was also supported by the \textit{Qiushi Academic - Dongliang} Talent Cultivation Program of Renmin University of China (Grant No. RUC24QSDL040). Computational resources were provided by Physical Laboratory of High Performance Computing in Renmin University of China. 
\end{acknowledgments}

\appendix

\section{Method: dynamical mean-field theory \label{sec:DMFT}}

The dynamical mean-field theory (DMFT) \cite{Georges1996rmp} helps us understand electron behavior in strongly correlated materials in a non-perturbative way. The DMFT neglects spatial correlations (assuming the self-energy of a system is local) and is exact in three limits: (1) non-interacting, (2) atomic (infinite interactions), or (3) infinite dimensions.

As the DMFT disregards spatial correlations and focuses solely on local correlations, it maps the lattice model (\ref{eq:lattmodel}) into a quantum impurity model
\begin{equation}\label{eq:impmodel}
  H_{\rm imp}=H_{\text{loc}}+H_{\text{bath}}+H_{\text{hyb}},
\end{equation}
where $H_{\rm{imp}}$ is the Hamiltonian, and is determined self-consistently. $H_{\rm{loc}}$ is the local part of the lattice Hamiltonian. $H_{\rm{bath}}$ describes an electronic bath for the impurity. And $H_{\rm{hyb}}$ represents the hybridization between the impurity and the bath.
\begin{align}
  H_{\rm loc}&=U\sum_{l}(n_{l\uparrow}-\frac{1}{2})(n_{l\downarrow}-\frac{1}{2}) + U^\prime(n_{1}-1)(n_{2}-1), \nonumber\\
  H_{\rm bath}&=\sum_{lk\sigma}\epsilon_{lk}b_{lk\sigma}^{\dagger}b_{lk\sigma},  \label{eq:impmodel_expanded}\\
  H_{\rm hyb}&=\sum_{lk\sigma}V_{lk}c_{l\sigma}^{\dagger}b_{lk\sigma}+h.c. ,\nonumber
\end{align}
where $b_{lk\sigma}^{\dagger}$ creates an electron with spin $\sigma$ at bath site $lk$; $n_{l\sigma}$ is the number operator; $\epsilon_{lk}$ is the onsite energy of the bath site $lk$ for the impurity orbital $l$; $V_{lk}$ is the hybridization strength between the impurity orbital $l$ and its corresponding bath site $lk$.
$H_{\rm{bath}}$ and $H_{\rm{hyb}}$ are determined by requiring $G_{\rm{imp, 0}}^{-1} = \mathcal{G}^{-1}$, where $G_{\rm{imp, 0}}$ is the non-interacting impurity Green's function and $\mathcal{G}$ is the Weiss field for a lattice site. Here, the bath is discretized, i.e., $k = 1, 2, ..., n_{\rm b}$, where $n_{\rm b}$ is a {\em finite} number denoting the number of bath sites for each impurity orbital $l$. 
$\epsilon_{lk}$ and $V_{lk}$ are determined by fitting the impurity hybridization function $\varGamma_{\rm imp}(z)$ to the local hybridization function $\varGamma_{{\rm loc},l}(z) = z - \mathcal{G}_l^{-1}(z)$, where $z$ is the complex frequency. And the impurity hybridization function is expressed as:
 
\begin{equation}\label{eq:hyb_imp}
  \varGamma_{{\rm imp},l}(z) = \sum_{k}\frac{|V_{lk}|^{2}}{z-\epsilon_{lk}}.
\end{equation}

The DMFT self-consistency equations \cite{Georges1996rmp} are then obtained by assuming that the single-site impurity model can captures the local dynamics of the original lattice model, i.e., by requiring the Green's function of the quantum impurity model equal to the local Green's function of the original lattice model, $G_{{\rm imp}}(z) = G_{{\rm loc}}(z)$, with identical self-energies.

Given an infinite coordination number of the Bethe lattice, the DMFT becomes an exact theory for this model, resulting in a notably simple form for the local Green's function \cite{Georges1996rmp, Nunez2018prb} of the lattice Hamiltonian (\ref{eq:lattmodel}),
\begin{equation}\label{eq:lattice_locG}
  G_{{\rm loc},l}(z) = \frac{z-\Sigma_{{\rm loc}, l}(z) - \sqrt{(z-\Sigma_{{\rm loc}, l}(z))^2 - 4t_{l}^2}}{2t_{l}^2},
\end{equation}

In the DMFT framework, the self-consistency equations are solved iteratively as follows. Starting from an initial guess for the local self-energy $ \Sigma_{{\rm loc}}(z)$, we first compute $ G_{{\rm loc}}(z)$ using Eq.~(\ref{eq:lattice_locG}). The Weiss field is then obtained as $\mathcal{G}_{l}(z)= G_{{\rm loc}, l}^{-1} (z)+\Sigma _{{\rm loc}, l} (z)$. Using the Weiss field, we recalculate the local hybridization function. Next, we construct a new impurity Hamiltonian $H_{\text{imp}}$ (\ref{eq:impmodel}) with a hybridization function $\varGamma_{{\rm imp}}(z)$ (\ref{eq:hyb_imp}) determined by fitting $\varGamma_{{\rm loc}}(z)$. We then solve the impurity model using an appropriate impurity solver to obtain a new impurity self-energy $\Sigma_{{\rm imp}}(z)$. Updating lattice local self-energy $\Sigma_{{\rm loc}}(z)$ with $\Sigma_{{\rm imp}}(z)$, we recalculate $G_{{\rm loc}}(z)$ and repeat the entire procedure until $G_{{\rm imp}}(z) = G_{{\rm loc}}(z)$ is satisfied within an acceptable error tolerance.

\section{Impurity solver: natural orbitals renormalization group\label{sec:NORG}}

Solving a quantum impurity model in practice is challenging, as it is also a strongly correlated system \cite{Georges1996rmp}. However, the correlation in a quantum impurity model differs significantly from that in a regular strongly correlated system, since the impurity orbitals can only entangle with a finite number of degrees of freedom in the bath. We refer to this property of a quantum impurity model as sparse correlation \cite{He2014prb}. Being underlied by it, He and Lu proposed the natural orbitals renormalization group (NORG) \cite{He2014prb, He2015prb, Zheng2018cpl} to find the ground state of a quantum impurity model, which selects many-body basis states based on the eigenvalues of the single-particle density matrix $D$ of the ground state, with matrix elements defined as
\begin{equation}\label{eq:density_matrix}
  D_{\alpha \beta} = \langle \psi_{\rm gs} | c_\alpha^\dagger c_\beta |\psi_{\rm gs} \rangle,
\end{equation}
where $c_\alpha^\dagger$ creates an electron at orbital $\alpha$, and $\psi_{\rm gs}$ is the ground state wave function. The eigenvectors of $D$ define the nature orbitals for the ground state. The occupancy numbers of the natural orbitals are the corresponding eigenvalues of the density matrix $D$. The ground state $|\psi_{\rm gs}\rangle$, when expanded in the natural orbital basis, consists of very few Slater determinants compared to the size of the entire Hilbert space \cite{He2014prb,Debertolis2021prb}.

The NORG has been demonstrated as a powerful method for solving quantum impurity models, as it can efficiently and explicitly find the ground state and accurately calculate the Green's functions for a multi-impurity/orbital Anderson model \cite{He2014prb}. By using it on a two-impurity Kondo problem with up to 1022 bath sites, a long-standing discrepancy between the NRG and quantum Monte Carlo studies has been resolved \cite{He2015prb}. Recently, we have applied NORG as an impurity solver for DMFT to study the hight-temperature superconductivity and electron correlation in $\rm La_3Ni_2O_7$ \cite{Tian2024prb,Chen2024arxiv}.

\section{Bath fitting}\label{sec:bathfit}

\begin{figure}[htbp!]
  \includegraphics[width=8.6cm]{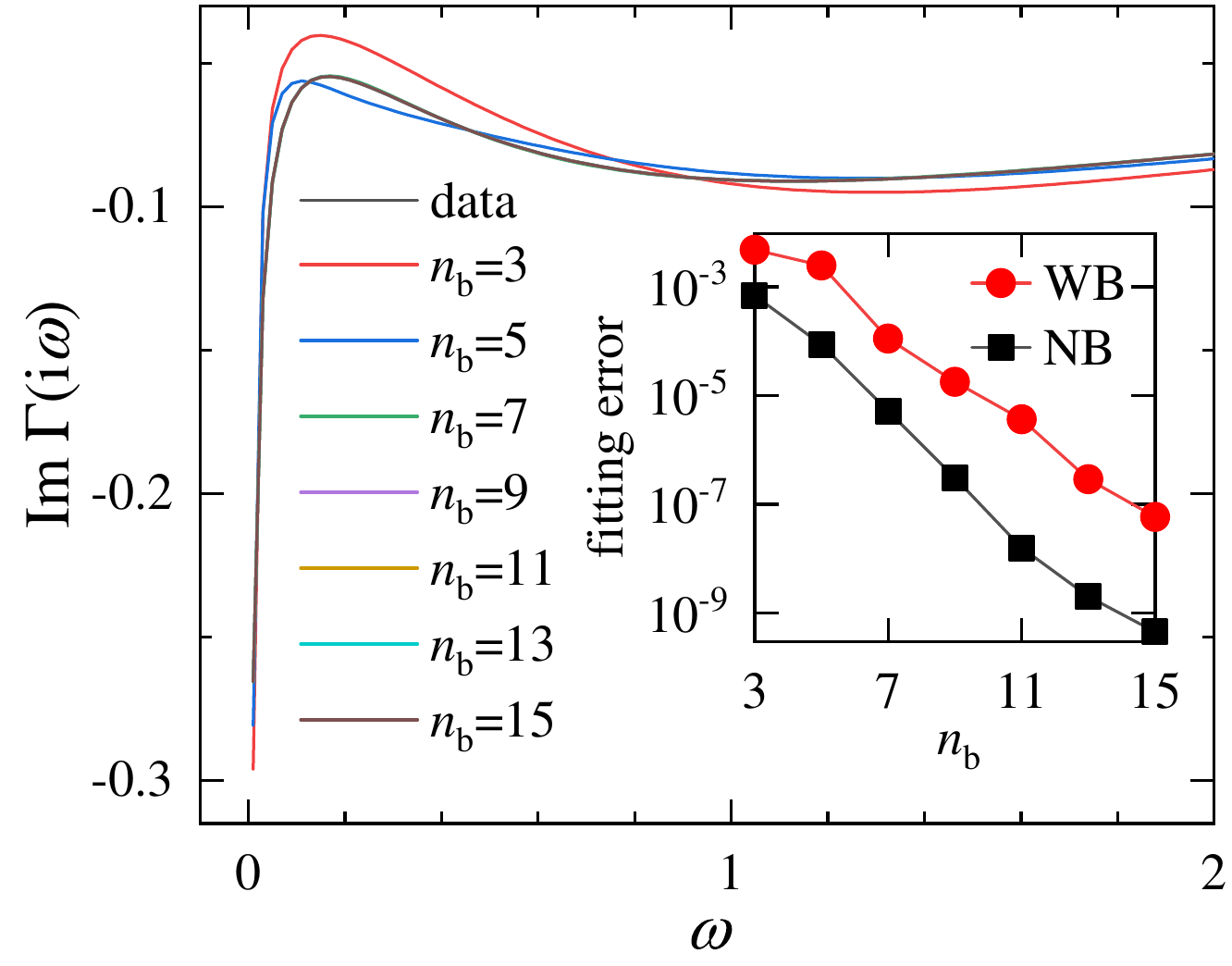}
  \caption{
    An example of the bath fitting. The imaginary part of the impurity hybridization function $\Gamma_{\rm imp}(i \omega)$ (\ref{eq:hyb_imp}) for the WB is shown. $U = 3$, $\Delta = 0.3$, and $t_2 = 0.5 t_1$. The inset shows the fitting error for the two bands with different $n_{\rm b}$.
    }\label{fig_FigS1}
\end{figure}

$H_{\rm{bath}}$ and $H_{\rm{hyb}}$ in the quantum impurity model (\ref{eq:impmodel_expanded}) are determined by fitting $i\omega - \mathcal{G}^{-1}(i \omega)$ to the impurity hybridization function $\Gamma_{\rm imp}(i \omega)$ (\ref{eq:hyb_imp}). To assess the accuracy of our fitting, we calculated the fitting errors with different $n_{\rm b}$, which denotes the number of discretized bath sites per impurity orbital. An example of the fitting is shown in Fig.~\ref{fig_FigS1}. The the fitting quality improves as $n_{\rm b}$ increases. As shown in the inset of Fig.~\ref{fig_FigS1}, the fitting error decreases exponentially as $n_{\rm b}$ increases. Notably, when $n_{\rm b} \geqslant 7$, the error is reduced to be smaller than $10^{-4}$ for both the wide band (WB) and narrow band (NB).

\section{Some details about the spectra\label{sec:IBK_delta_energy}}

\begin{figure}[htbp!]
    \includegraphics[width=8.6cm]{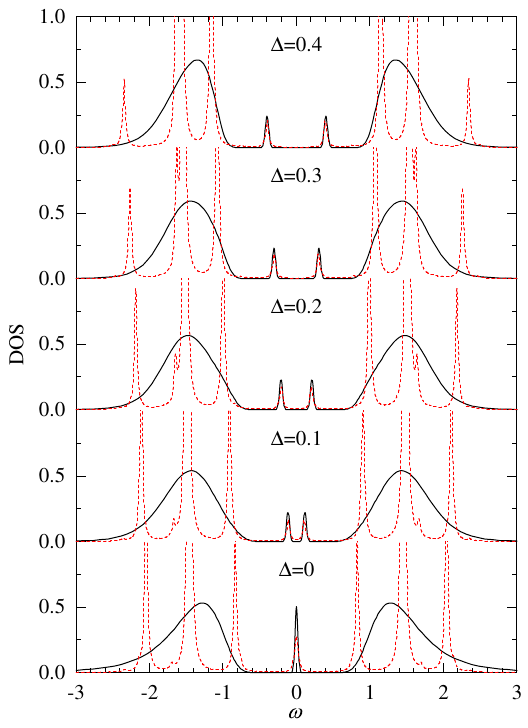}
    \caption{
      DOS of the narrow band for $\Delta = 0, 0.1, ..., 0.4$ (with $U = 3$ and $t_2 = 0.1 t_1$). The black solid lines indicate the results obtained through numerical analytical continuation using the maximum entropy method, while the red dashed lines indicate the results calculated directly on the real-frequency axis using the Lanczos method implemented with the NORG.
    }\label{fig_FigS2}
  \end{figure}
  
  We compare the real-frequency local spectra calculated respectively using the Lanczos and maximum entropy methods for the same quantum impurity model with $\Delta = 0, 0.1, ..., 0.4$, $U = 3$, and $t_2 = 0.1 t_1$. The Lanczos method directly obtains the real-frequency spectra \cite{Jakli1994prb, Georges1996rmp, Weisse2006rmp, Niu2019prb}. The method is readily integrated into the NORG. The maximum entropy method obtains the real-frequency spectra from the Matsubara Green's function through a numerical analytical continuation. For this method, the \emph{ana\_cont} package \cite{Kaufmann2023cpc} is employed. In Fig.~\ref{fig_FigS2}, we present the DOSs of the NB with different $\Delta \equiv U - U'$. The DOSs obtained via both the methods match very well in the low-frequency region, particularly for the IBK QP peaks. The IBK peaks are located at $\omega = \pm \Delta$, which is consistent with the results in Ref. \cite{Nunez2018prb}. In the high-frequency region, the results from the Lanczos method show discrete peaks due to the bath discretization.

  \section{Absence of orbital selective Mott transition\label{sec:absOSMT}}

\begin{figure}[H]
    \includegraphics[width=8.6cm]{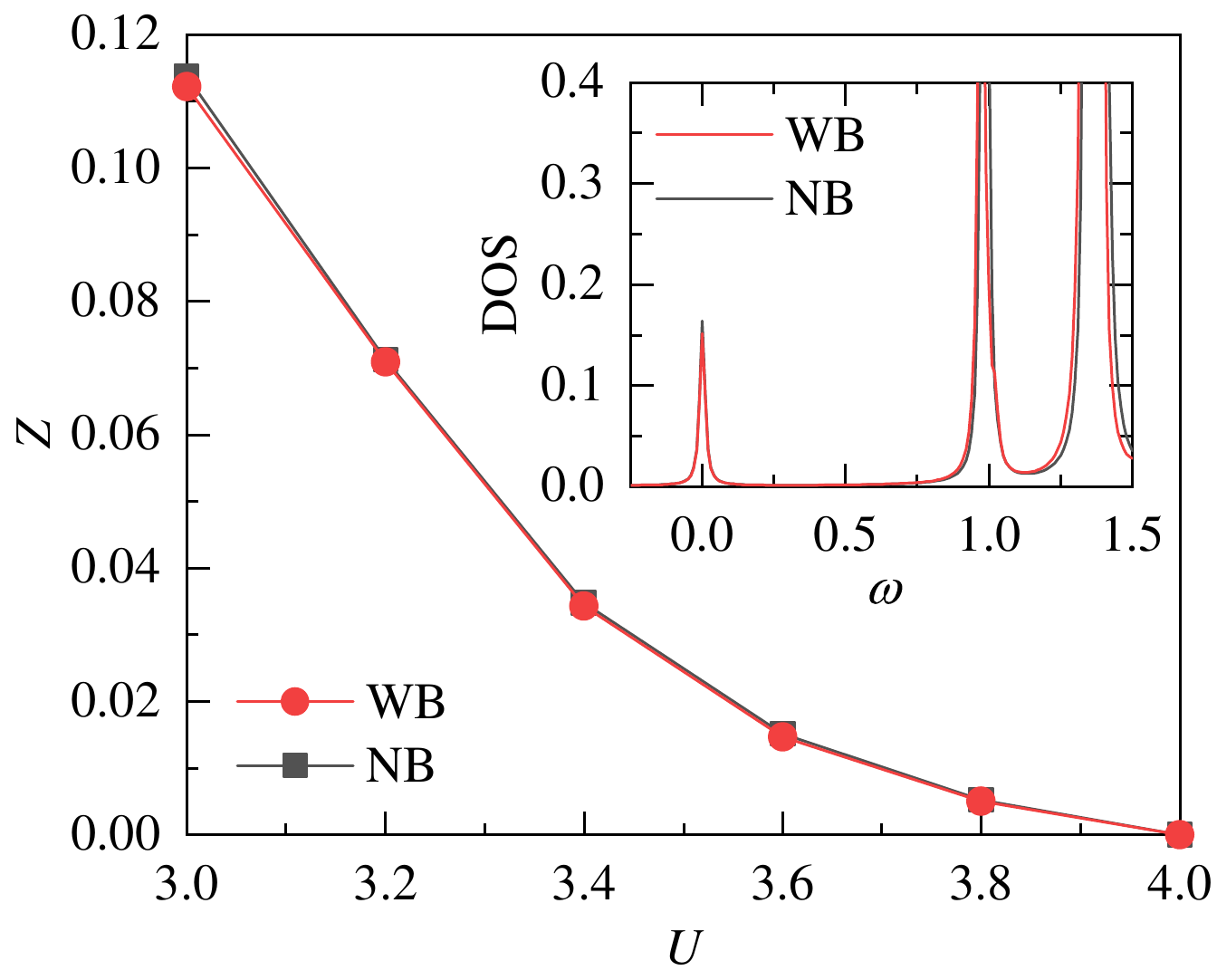}
    \caption{
      Quasiparticle weight $Z$ as a function of $U$. $\Delta = 0$ and $t_2 = 0.5 t_1$. The inset shows the DOS for $U = 3.8$.
      }\label{fig_FigS3}
  \end{figure}

  As discussed in Sec.~\ref{sec:IBK_delta_energy} and the main text, the IBK QP peaks are located at $\omega = \pm \Delta$. When $\Delta = 0$, the IBK peaks merge at zero frequency, as shown in Fig.~\ref{fig_FigS2}. Consequently, both the WB and NB exhibit central peaks: a Kondo-like QP peak in the WB and an IBK QP peak in the NB. Fig.~\ref{fig_FigS3} shows the quasiparticle weight $Z$ as a function of $U$ with $\Delta = 0$ and $t_2 = 0.5 t_1$, where $Z^{-1}=1-\left.\frac{\partial \operatorname{Re} \Sigma(\omega)}{\partial \omega}\right|_{\omega=0}$. The quasiparticle weights of the WB and NB remain equal across different values of $U$. The inset of Fig.~\ref{fig_FigS3} shows that the QP peaks of the WB and NB overlap around $\omega = 0$, indicating that the two peaks have the same spectral weight. The quasiparticle weights decrease as $U$ increases, and eventually vanish at $U = 4.0$. This indicates that these peaks disappear simultaneously, demonstrating the absence of an orbital-selective Mott transition. Similar results were reported in Refs. \cite{Rong2017prb, Komijani2017prb, Nunez2018prb}.
  
\bibliography{ibkbs}

\end{document}